\documentclass[%
 reprint,
 amsmath,amssymb,
 aps,
prb,
]{revtex4-1}
\usepackage{float} 
\usepackage{graphicx}
\usepackage{dcolumn}
\usepackage{bm}
\usepackage{hyperref}
\usepackage[mathlines]{lineno}
\usepackage{mathtools, cuted}
\usepackage[usenames,dvipsnames,svgnames,table]{xcolor}

\begin{document}

\preprint{APS/123-QED}
\title{Enhancement of recoil optical forces via high-$\mathbf{k}$ plasmons on thin metallic films}

\author{ J. A. Gir\'on-Sedas}
 \affiliation{%
 Departamento de F\'isica, Universidad del Valle, AA 25360, Cali, Colombia }%
\author{Francisco J. Rodr\'iguez-Fortu\~no}%
 \email{francisco.rodriguez$_$fortuno@kcl.ac.uk }
\affiliation{%
 *Department of Physics and London Centre for Nanotechnology, King's College London, WC2R 2LS, United Kingdom}%

\date{\today}

\begin{abstract}
The recoil optical force that acts on emitters near a surface or waveguide relies on near-field directionality and conservation of momentum. It features desirable properties uncommon in optical forces, such as the ability to produce it via wide-area illumination of vast numbers of particles without the need for focusing, or being dynamically switchable via the polarization of light. Unfortunately, these recoil forces are usually very weak and have not been experimentally observed in small dipolar particles. Some works theoretically demonstrate orders-of-magnitude enhancement of these forces via complex nano-structuring involving hyperbolic surfaces or metamaterials, complicating the fabrication and experimental demonstration. In this work we theoretically and numerically show enhancement of the lateral recoil force by simply using thin metallic films, which support ultra-high-momentum plasmonic modes. The high-momentum carried by these modes impart a correspondingly large recoil force on the dipole, enhancing the force by several orders of magnitude in a remarkably simple geometry, bringing it closer to practical applications.

\end{abstract}

\pacs{Valid PACS appear here}
\maketitle


Optical forces, enabling the manipulation of nanoparticles through light-matter interactions, have allowed important progress in many areas ranging from ultra-cold matter physics to biology\cite{Ashkin1970,Ashkin1971,Ashkin1986,Bagnato1987,Chu1998,Phillips1998,Ashkin2000,Grier2003,Allen2003,Loudon2012}. Most optical forces rely on optical traps, gradients of field intensity created via focusing, ideal for the precise manipulation of individual nanoparticles, but challenging to adapt to massively-parallel manipulation of several particles simultaneously. An alternative approach that overcomes this obstacle is to rely on the scattering of particles under non-focused plane-wave-like illumination with no intensity gradients. Because light carries linear momentum, it follows that if a particle scatters light in a given direction, it must experience a mechanical recoil pushing it in the opposite direction, due to conservation of momentum. This has been exploited in free-standing particles for applications such as tractor beams \cite{Chen2011}, and more recently in particles placed near surfaces or waveguides\cite{Kingsley-Smith2018,Petrov2016,Fortuno2013,Petersen2014,OConnor2014,Fortuno2015,Wang2014,Sukhov2015,Ivinskaya2017}. When a particle is illuminated and there is a waveguide or surface nearby, the near-field scattering fields can couple into the guided modes of the waveguide or surface modes such as surface plasmons, which propagate away from the particle and hence impart a corresponding mechanical recoil to the particle \cite{Kingsley-Smith2018,Petrov2016}. This mechanism does not rely on gradients of the illuminating light, and so it doesn't require focusing and occurs under plane wave illumination in a wide area. The directionality of the near-field excitation, required to achieve a net force, can be achieved by controlling the electric and/or magnetic polarization of the scatterer \cite{Kingsley-Smith2018,Petrov2016}. Interestingly, these optical forces can be controlled using optical degrees of freedom, such as the polarization of light. Thanks to spin-orbit interaction of light, the polarization state of light may modify and control the spatial degrees of freedom of light, i.e. its propagation direction and scattering \cite{Bliokh2015a}. In particular, it is well-known that a circularly polarized dipole or scatterer near a waveguide will excite waveguide modes directionally in a lateral direction \cite{Fortuno2013,Petersen2014,OConnor2014} and hence this will result in a polarization-dependent recoil lateral optical force, whose direction is switchable with the polarization handedness of the illumination \cite{Fortuno2015,Wang2014,Sukhov2015}, thus enabling a simple non-mechanical method for dynamically controlling the direction of the force. Alternatively to plane wave illumination, the polarization pattern of tightly focused linearly polarized beams can exhibit spinning polarizations away from the beam's axis, which can be used to enable optical trapping or anti-trapping near surfaces based on the same recoil-force principle\cite{Ivinskaya2017}, instead of relying on illuminating field gradients. The same recoil optical forces also lead to lateral Casimir forces on rotating particles near smooth surfaces \cite{Manjavacas2017}.

Despite much theoretical works, these polarization-controlled recoil lateral forces are very weak under normal circumstances, and their experimental measurement has only been achieved on large 4.5 $\mu$m particles at wavelengths of 532 nm \cite{Sukhov2015}, well outside the dipole approximation, and on macroscopic birefringent objects \cite{Magallanes2018}. The lateral force due to near-field directionality of a circularly polarized scatterer in the dipole approximation, which would have vast applications on optical manipulation of nanoparticles and molecules, has proven too weak for an experimental demonstration to date. However, some works have theoretically proposed methods to greatly enhance this force. One approach has been to use hyperbolic metasurfaces, theoretically predicting an increase on the lateral force by roughly three orders of magnitude \cite{Paul2019PRB}. Hyperbolic metasurfaces are planar metamaterials which show an anisotropic surface conductivity \cite{Liu2013,Yu2014,Kildishev2013,Yermakov2015} and were experimentally demonstrated using single-crystal silver nanostructures \cite{High2015}, but such materials are challenging to fabricate in large areas useful for measurement and application of lateral optical forces, with no experimental demonstration yet produced of the enhanced force. 

Another approach to achieve enhanced lateral forces from circularly polarized dipoles was theoretically predicted with the use of bulk hyperbolic metamaterials \cite{Ivinskaya2018}. This is different to hyperbolic metasurfaces because, rather than requiring an anisotropic surface conductivity, it requires an anisotropic bulk permittivity. Such hyperbolic metamaterials are arguably easier to produce in large centimeter-sized areas via the use of metallic nanorods or alternating metal-dielectric thin layers \cite{Nicholls2017,Kabashin2009,Simovski2012,Poddubny2013}. Circularly polarized dipoles are known to exhibit directionality near hyperbolic metamaterials\cite{Kapitanova2014}, where instead of exciting surface plasmons or single guided modes, they excite combinations of high-wavevector bulk modes forming subwavelength `rays' inside the hyperbolic metamaterial, but still experience the associated lateral recoil force. Yet no experimental demonstration of the enhancement has been achieved in this case either. We believe that a simpler geometry is desirable.

The key feature, common in all the approaches for enhancement of recoil lateral forces, is the existence of high-wavevector (high-$\mathbf{k}$) modes in the surface or in the bulk. Such modes, excited by the dipole, possess a huge phase gradient due to the very small wavelength of the mode, which manifests as a strong recoil force acting back on the dipole. In terms of momentum conservation, the individual photons or plasmons of the excited modes have a very large momentum $\mathbf{p} = \hbar \mathbf{k}$ owing to the large value of $\mathbf{k}$, and hence each produces a strong recoil `kick' on the particle. This is a similar idea to the concept of a `super-kick' acting on particles placed near vortex beam singularities, in regions where the optical phase gradient is also huge \cite{Barnett2013}. Therefore, for a practical recoil lateral force enhancement, it would be advantageous to find the simplest structures, easy to reproduce in the laboratory and commercially, that support modes with very high wave-vector $\mathbf{k}$. The directional excitation of such modes would then result in an enhanced lateral recoil force. In this work we propose the use of a very thin metallic slab, known to support short-range surface plasmon with a dramatically reduced wavelength and correspondingly high $\mathbf{k}$ \cite{Burke1986,Berini2001,Barnes2003,Berini2009}. We theoretically and numerically prove several orders of magnitude enhancement of the lateral recoil force acting on circularly polarized emmiters near such thin films.


It is well known that a planar interface between a metal  with relative permittivity $\varepsilon_2$ and a dielectric with permittivity $\varepsilon_1$ supports surface plasmon polariton modes (SPPs) with a wave-vector given by $k_{\mathrm{SPP}}=k_0(\varepsilon_1 \varepsilon_2/(\varepsilon_1+\varepsilon_2))^{1/2}$. If a thick metal slab is sandwiched between two dielectrics $\varepsilon_{1}$ and $\varepsilon_{3}$, then, as long as the metal slab is thicker than the plasmon penetration depth,  two surface modes will exist independently in the two interfaces, as shown in Fig.\ \ref{figdispersion}. However, if the slab is thinner than the plasmon penetration depth $t \sim (1/2)(k_{\mathrm{SPP}}^2 - \varepsilon_2 k_0^2)^{-1/2}$, then coupling and mode splitting will occur between the two modes, forming two supermodes with even and odd symmetry in different field components, commonly known as the long-range (LR) and short-range (SR) surface plasmons \cite{Burke1986,Berini2001,Barnes2003,Berini2009}, depicted in Fig.\ \ref{figdispersion}. The SR-SPP is highly confined in the metal film and is ignored in many plasmonic applications due to its high losses and low propagation length -- hence its name, however, these are not limiting factors for the recoil forces that these modes may cause. The thinner the metal film, the stronger the splitting between the two modes, and the higher the wave-vector up to which the SR-SPP is pushed into, as shown in Fig.\ \ref{figdispersion}, resulting in extremely reduced modal wavelengths. In this work, as an example, we consider a gold film between free space and a $\mathrm{SiO_2}$ substrate. Both materials are widely used in the development of both photonic and plasmonic structures, operating in a broad frequency spectrum\cite{Malitson1965,Johnson1972}. We have used the free-electron Drude model fit \cite{Myroshnychenko2008} to the dielectric function of gold \cite{Johnson1972}.

\begin{figure}[ht]
\includegraphics[width=0.45\textwidth]{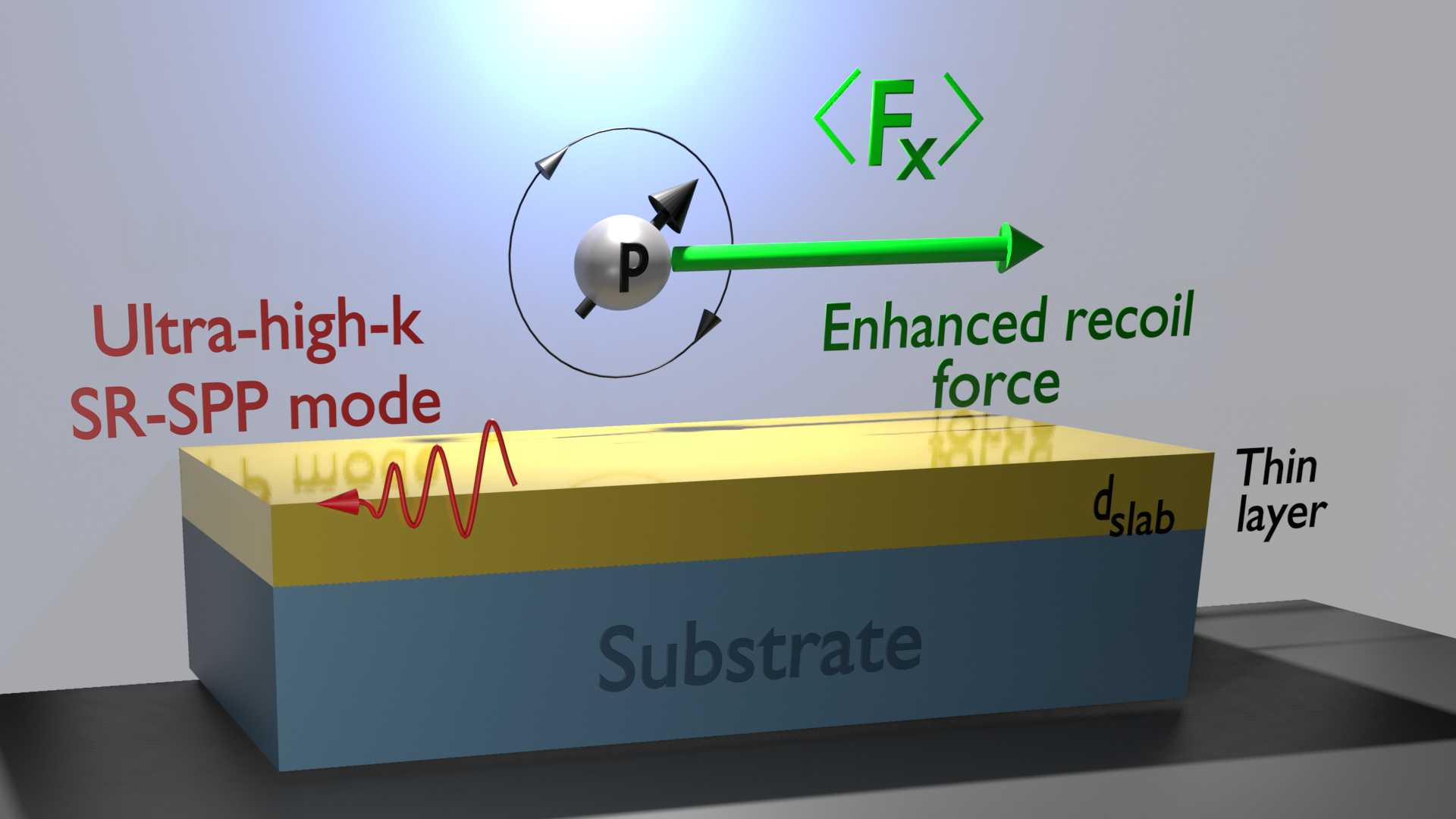}
\centering
\caption{\label{fig1}(Color online) Schematic diagram of a circular electric dipole above a thin metallic slab. }
\end{figure}

\begin{figure}[h]
\includegraphics[]{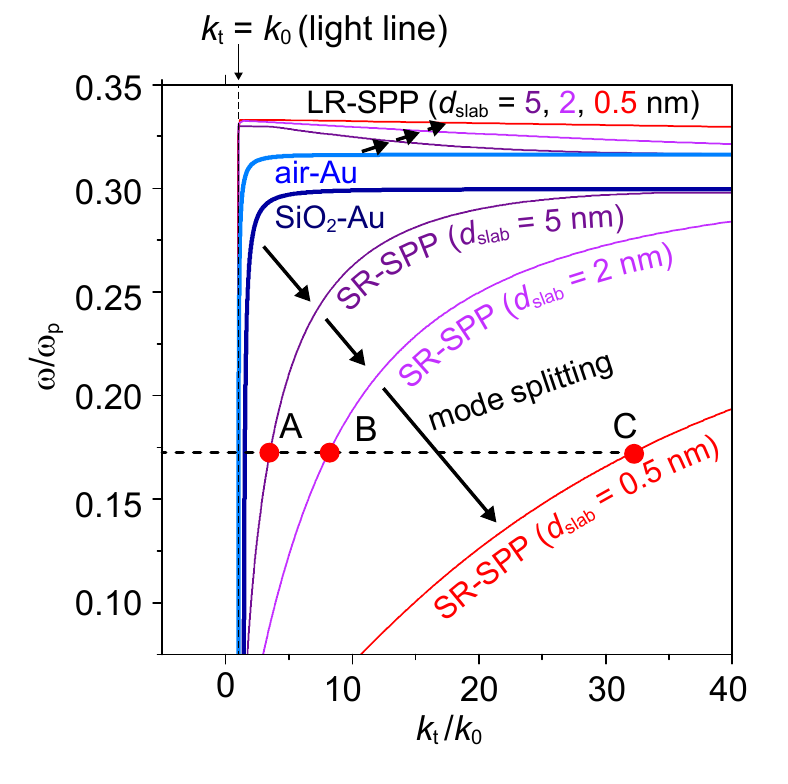}
\centering
\caption{\label{figdispersion}(Color online) Normalized dispersion curves of surface plasmon polariton modes propagating along a gold slab between air ($\varepsilon_1=1$) and fused silica ($\varepsilon_3=2.11$) for three thicknesses $d_{\mathrm{slab}}$ = 0.5 nm, 1 nm and 5 nm. The corresponding dispersion relation for single interface air-gold and $\mathrm{SiO_{2}}$-gold are presented. We used the dielectric function of gold as a free-electron Drude model  $\varepsilon_2=\varepsilon_{\infty}-\omega_p^2/(\omega^2+i \omega \tau^{-1} )$, where $\omega_p=9$ eV$/\hbar$, $\tau^{-1}=0.05$ eV$/\hbar$, and $\varepsilon_\infty=9$. The Drude model fits experimental gold data well in the infrared and red part of the spectrum. The locations A, B, C correspond to the SR-SPP mode at a wavelength $\lambda = 800$ nm.}
\end{figure}

Knowing that we can achieve high-$\mathbf{k}$ modes in thin metallic films, we now study how these affect the recoil optical forces that act on a nearby circularly polarized dipole, which in practice can represent any polarizable particle illuminated with polarized light. Our results are obtained analytically via the exact Green’s function formalism of a dipole over a surface; using the dipole angular spectrum approach\cite{Mandel1995,Nieto2006,Novotny2012,Picardi2017,Kingsley-Smith2018}. Consider an electric dipole source $\textbf{p}=(p_x,p_y,p_z)$ located at $\textbf{r}_0=(0,0,h)$, above a metallic slab of thickness $d_{\mathrm{slab}}$, whose surfaces are $z=0$ and $z=-d_{\mathrm{slab}}$, which has been grown on a dielectric substrate, as seen in Fig.\ \ref{fig1}. The dipole is radiating with an angular frequency $\omega$. The time-averaged optical force $\left\langle \textbf{F} \right\rangle$ acting on the dipole can be deduced from first principles --the Lorentz electromagnetic force acting on the oscillating charges of a dipole due to the backscattered fields from the surface-- and is well-known \cite{Gordon1980,Chaumet2000,Novotny2012}. It is given by $\langle \textbf{F} \rangle = \sum_{i=x,y,z} \frac{1}{2} \mathfrak{Re} \{p^{*}_{i} \nabla E_{i}  \}$, where $E_{x,y,z}$ represents the total electric field minus the self-fields of the dipole, and therefore includes any back-scattered fields or excited modes on the surface responsible for the recoil force \cite{Kingsley-Smith2018}.

Here, we will focus on the $x$ component of the force that acts over a polarized particle, which following previous works\cite{Fortuno2015}, after some mathematical steps and the only assumption that $\varepsilon_1$ is real (lossless upper medium), can be written in a compact manner as:

\begin{equation}
\label{fx}
\begin{split}
\left\langle F_x \right\rangle= P_{\mathrm{rad}}^{xz} \sigma_y \frac{3}{4 c_0 n_1^3}   \mathfrak{Im} \bigg\lbrace & \int \mathrm{d} k_{tr} e^{i 2 k_{z1}h} k_{tr}^3  r_{pp} \bigg\rbrace,
\end{split}
\end{equation}

\noindent where the integration is performed over the normalized transverse wave-vector, $k_{tr}=\frac{k_t}{k_0} \in [0,\infty]$, $P_{\mathrm{rad}}^{xz}=\frac{n_1^3 \omega^4 \left( |p_{x}|^2+|p_{z}|^2 \right)}{12\pi \varepsilon_0 \varepsilon_1 c_0^3}$ is the power radiated by the $x$ and $z$  components of the dipole, $n_1 = \sqrt{\varepsilon_1}$ is the refractive index, $\sigma_y=- \frac{2 \mathfrak{Im}[p_x^* p_z]}{|p_{x}|^2+|p_{z}|^2} \in [0,1]$ is the dipole ‘spin’ along the $y$-axis, $k_{z1} = k_0(n_1^2-k_{tr}^2)^{1/2}$ is the z-component of the wave-vector, $k_0 = 2\pi/\lambda$ is the wavenumber of free space, $\lambda$ is the wavelength, and $r_{pp}$ is the well-known Fresnel reflection coefficient for p-polarization on a slab, which also has a dependency with the transverse wave-vector and the wavelength $r_{pp}(k_{tr},\lambda)$. As is well known\cite{Fortuno2015}, a lateral force in the $x$-direction appears when the dipole has a spin along the $y$-direction, i.e. spinning in the $xz$ plane. In the following results, we will consider the case that achieves the maximum force along $+x$: a clockwise circular dipole $\mathbf{p}=(1,0,-i)$, with $\sigma_y=1$, which excites modes directionally in the $-x$ direction and experiences the corresponding recoil directed along $+x$.

\begin{figure}[ht]
\includegraphics[]{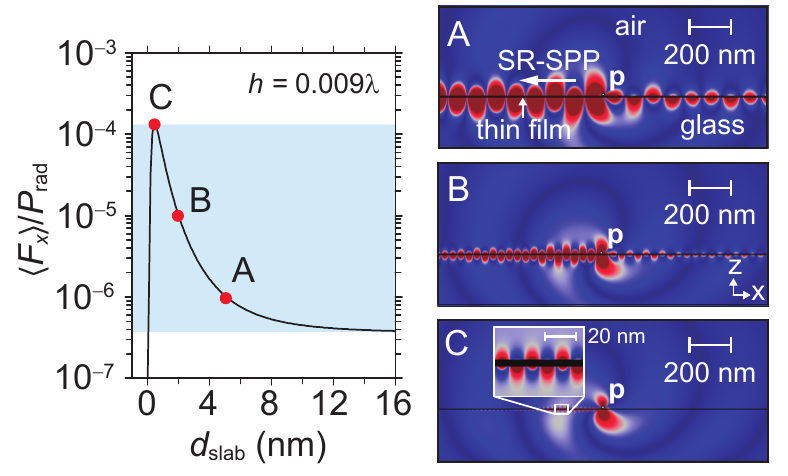}
\centering
\caption{\label{fig2}(Color online) Left: Thickness dependence of the time-averaged normalized lateral force when a circularly polarized dipole, $\mathbf{p}=(1,0,-i)$, is emitting with a wavelength radiation of $\lambda=800$ nm over a gold slab. Right:  COMSOL simulation of the magnetic field distribution ($H_y$) surrounding the dipole for three different slab thicknesses (A) 5 nm, (B) 1 nm, (C) 0.5 nm, corresponding to the dispersion points in Fig.\ \ref{figdispersion}. The dipole is located at a distance $h=0.009\lambda$ above the gold slab.}
\end{figure}

Let us calculate the lateral force when the dipole is placed near thin metallic slabs. As one can see in Eq.\ \ref{fx}, the high-$\mathbf{k}$ modes in the film will manifest themselves as a peak in $r_{pp}$ at a very high value of $k_{tr} = k_{\mathrm{SPP}}/k_0$ and will be weighted by $k_{tr}^3$, greatly enhancing the lateral force. Let us consider the same thin films as shown in Fig.\ \ref{figdispersion}. The exact calculation of the force $\left\langle F_{x} \right\rangle$ via Eq.\ \ref{fx} is shown in Fig.\ \ref{fig2}, for varying values of the metal thickness $d_{\mathrm{slab}}$. The polarization of the dipole is $\mathbf{p}=(1,0,-i)$ and it is located at a subwavelength distance $h=0.009 \lambda$ over the metallic slab. We notice that the force experiences a dramatic increase of nearly three orders of magnitude when the thickness of the slab is very thin, reaching an optimal strength around 0.5 nm. The figure insets also show the magnetic field distribution ($H_y$) for three values of the thickness. The directional excitation of guided modes to the left is observed in each one of these cases, resulting in recoil, but there is a clear distinctive feature: the wavelength of the plasmon decreases dramatically for the thinner films. This is as expected due to the short-range surface plasmon having an ultra high $k$ value \cite{Burke1986,Berini2001,Barnes2003,Berini2009} as discussed earlier. Such a high $k$ and the associated small wavelength produces a huge phase gradient at the location of the dipole, resulting in the increased lateral recoil force, in accordance with our initial expectation.


As discussed earlier, this has a quantum interpretation: the amount of momentum carried by each quantized plasmon in the excited plasmonic mode $\mathbf{p}=\hbar \mathbf{k}$ is proportional to $\mathbf{k} = -k_{\mathrm{SPP}} \hat{\mathbf{x}}$, hence, for each excited plasmon with ultra-high momentum, the dipole source must fulfill conservation of momentum and experience a `super-kick' in the opposite direction, similar to the `super-kick' predicted for particles near phase singularities \cite{Barnett2013}. The fields in the three cases depicted in Fig.\ \ref{fig2} are oscillating at the same angular frequency $\omega$, hence the energy carried per quantum $E=\hbar \omega$ is the same, while the momentum is much higher for the thinnest film. For each energy packet transferred to the plasmon mode, the recoil `kick' experienced by the dipole is far greater in the high-$\mathbf{k}$ plasmonic mode.

One may ask what is the theoretical limit of this effect. Can we keep reducing the film to achieve infinitely higher wave-vectors? Unfortunately, there is an unavoidable limitation. The larger the propagation wave-vector of the mode becomes, the more confined it is to the interface, and the stronger the evanescent decay into the dielectric. This will reduce the coupling efficiency between the dipole and the mode. In the quantum picture, fewer energy packets will be excited by the dipole to produce the super-kick. The coupling is mathematically accounted for on Eq.\ \ref{fx} by the exponential decay term $e^{2ik_{z1}h} = e^{-2h k_0(k_{tr}^2-\varepsilon_2)^{1/2}}$ inside the integral which dampens the peak in $r_{pp}(k_{tr})$ that occurs at $k_{tr}=k_\mathrm{SPP}/k_0 \gg 1$ due to the SPP. This dampening trades off against the cubic enhancement of the term $k_{tr}^3$, resulting in the existence of an optimal thickness of the metallic layer, as clearly shown in Fig.\ \ref{fig2}. This optimum depends on the distance $h$ between the dipole and the surface. The greatest enhancements are achieved when the dipole is placed very close to the surface, therefore the technique will work best for atoms and molecules, which can be placed at distances of hundredths of a wavelength or less. This theoretical limitation applies to any proposal based on high-$\mathbf{k}$ modes, such as the existing proposals based on hyperbolic materials.

In view of an experiment, the fabrication is simpler than that of hyperbolic metamaterials and metasurfaces, but still faces some challenges. The enhancement of the force relies on the extremely small wavelength and high-confinement of the modes to the surface, hence the mode will be highly susceptible to surface roughness. The aim is, therefore, a very thin and smooth film. While in our calculations we used gold as the archetypal plasmonic material, the physics is identical for any plasmonic alternative, such as most metals, some semiconductors, conductive metal oxides such as indium-tin-oxide (ITO), and others\cite{Naik2013}. The choice of material should be decided by the ease of fabrication of smooth thin films and the desired wavelength of operation. Even the use of plasmonic 2D conductive materials such as doped graphene could be considered.

In conclusion, we have theoretically shown several orders of magnitude enhancement of the lateral recoil optical forces with the use of a simple system: thin plasmonic films supporting an ultra-high-$\mathbf{k}$ hybridized super-mode. The geometry of such system is simpler than hyperbolic metasurfaces or metamaterials, and is suitable for fabrication on wide areas, and hence this increases the prospects of observing these lateral forces experimentally and brings the force a step closer to applications such as the development of efficient platforms for optical manipulation of nanoparticles in optical `conveyor belts' relying on lateral forces, or manipulation and separation of chiral nanoparticles. The recoil force on circular dipoles is directly responsible for the lateral Casimir force acting on rotating particles near a smooth surface\cite{Manjavacas2017}, hence the enhancement presented here directly translates into a corresponding enhancement of lateral Casimir forces.

We acknowledge the financial support from the Colombian agency COLCIENCIAS (Postdoctoral stays - No. 811)  and the European Research Council Starting Grant ERC-2016-STG-714151-PSINFONI.

\bibliographystyle{apsrev4-1}

\end{document}